\documentclass{article}

\usepackage[preprint]{neurips_2026}

\usepackage{amsmath}
\usepackage{booktabs}
\usepackage{multirow}
\usepackage{makecell}
\usepackage{graphicx}
\usepackage{pifont}
\usepackage{tabularx}
\usepackage{wrapfig}
\usepackage{subcaption}
\newcommand{\cmark}{\ding{51}}
\newcommand{\xmark}{\ding{55}}

\usepackage[utf8]{inputenc} 
\usepackage[T1]{fontenc}    
\usepackage{hyperref}       
\usepackage{url}            
\usepackage{booktabs}       
\usepackage{amsfonts}       
\usepackage{nicefrac}       
\usepackage{microtype}      
\usepackage{xcolor}         

\title{Membership Inference Attacks on Vision-Language-Action Models}

\author{
Yuefeng Peng$^{1}$\thanks{Equal contribution.} \quad
Mingzhe Li$^{1}$\footnotemark[1] \quad
Kejing Xia$^{2}$ \quad
Renhao Zhang$^{1}$ \quad
Amir Houmansadr$^{1}$ \\
$^{1}$University of Massachusetts Amherst \quad
$^{2}$Georgia Institute of Technology
}

\begin{document}
\maketitle

\begin{abstract}
Membership inference attacks (MIAs) have been extensively studied in large language models (LLMs) and vision-language models (VLMs), yet their implications for vision-language-action (VLA) models remain largely unexplored. VLA models differ from standard LLMs and VLMs in several important ways: they are often fine-tuned for many epochs on relatively small embodied datasets, operate over constrained and structured action spaces, and expose action outputs that can be observed as executable behaviors and temporally correlated trajectories. These characteristics suggest a distinct and potentially more informative attack surface for membership inference. In this work, we present the first systematic study of MIAs against VLA systems. We formalize two membership inference settings for VLA models: sample-level inference over individual transition samples and trajectory-level inference over complete embodied demonstrations. We further develop a suite of attack methods under multiple access regimes, including strict black-box access. Our attacks exploit both classic MIA signals, such as token likelihood, and VLA-specific signals, such as observable action errors and temporal motion patterns. Across multiple VLA benchmarks and representative VLA models, these attacks achieve strong inference performance, showing that VLA models are highly vulnerable to membership inference. Notably, black-box attacks based only on generated actions achieve strong performance, highlighting a practical privacy risk for deployed embodied AI systems. Our findings reveal a previously underexplored privacy risk in robotic and embodied AI, and underscore the need for dedicated privacy evaluation and defenses for VLA models.
\end{abstract}

\section{Introduction}

Recent advances in multimodal learning have led to the emergence of vision-language-action (VLA) models~\cite{rt12022arxiv,rt22023arxiv,black2024pi_0,black2025pi_}, which map visual and textual inputs to executable actions in embodied environments such as robotics and interactive agents. By producing control signals or action sequences, VLA systems enable seamless perception-to-action pipelines and have shown strong performance across a range of real-world tasks~\cite{rt12022arxiv,rt22023arxiv,o2024open}. As these models are increasingly deployed in safety-critical and privacy-sensitive settings~\cite{tian2024drivevlm,fu2025orion,li2025robonurse}, understanding their privacy risks becomes essential.

A growing body of work has demonstrated that modern machine learning models are vulnerable to membership inference attacks (MIAs), where an adversary aims to determine whether a given data point was used during training~\cite{shokri2017membership,nasr2019comprehensive,carlini2022membership,peng2024oslo}. While MIAs have been extensively studied in large language models (LLMs)~\cite{duan2024membership,maini2024llm,fu2024membership,hayes2026exploring} and vision-language models (VLMs)~\cite{liu2026lomia,yin2026blackbox,hu2025membership,li2024membership}, their implications for VLA models remain largely unexplored. This gap is non-trivial: unlike LLMs and VLMs that primarily produce textual or semantic outputs, VLA models operate over structured action spaces and expose outputs as executable behaviors in physical or simulated environments. These differences raise a fundamental question: do VLA models introduce new avenues for MIAs?

This question is especially important in VLA systems, where models are trained on embodied datasets aggregated across users, environments, and institutions~\cite{o2024open,khazatsky2024droid}. Each training example corresponds to an interaction trajectory that may encode sensitive information such as private spaces, user behaviors, or task workflows. Membership inference thus provides a direct mechanism to determine whether specific embodied experiences---e.g., household routines~\cite{o2024open} or patient-care episodes~\cite{li2025robonurse}---have been absorbed into a deployed policy. Moreover, because embodied data is expensive to collect ~\cite{o2024open,liu2023libero} and often proprietary, membership inference can also be used to audit data provenance and detect unauthorized use of valuable datasets.

In this work, we take the first step toward answering this question by systematically studying MIAs in VLA models. Motivated by the structure of embodied data, we formalize two complementary inference settings: sample-level membership inference, which targets individual single-step decisions, and trajectory-level membership inference, which targets complete interaction sequences. For each setting, we develop a diverse set of attack methods, including both classical MIA approaches (e.g., loss- and likelihood-based signals)~\cite{yeom2018privacy,carlini2022membership,mattern2023membership} and VLA-specific methods that exploit action outputs and temporal dynamics.

\begin{figure*}[t]
    \centering
    \includegraphics[width=0.98\textwidth]{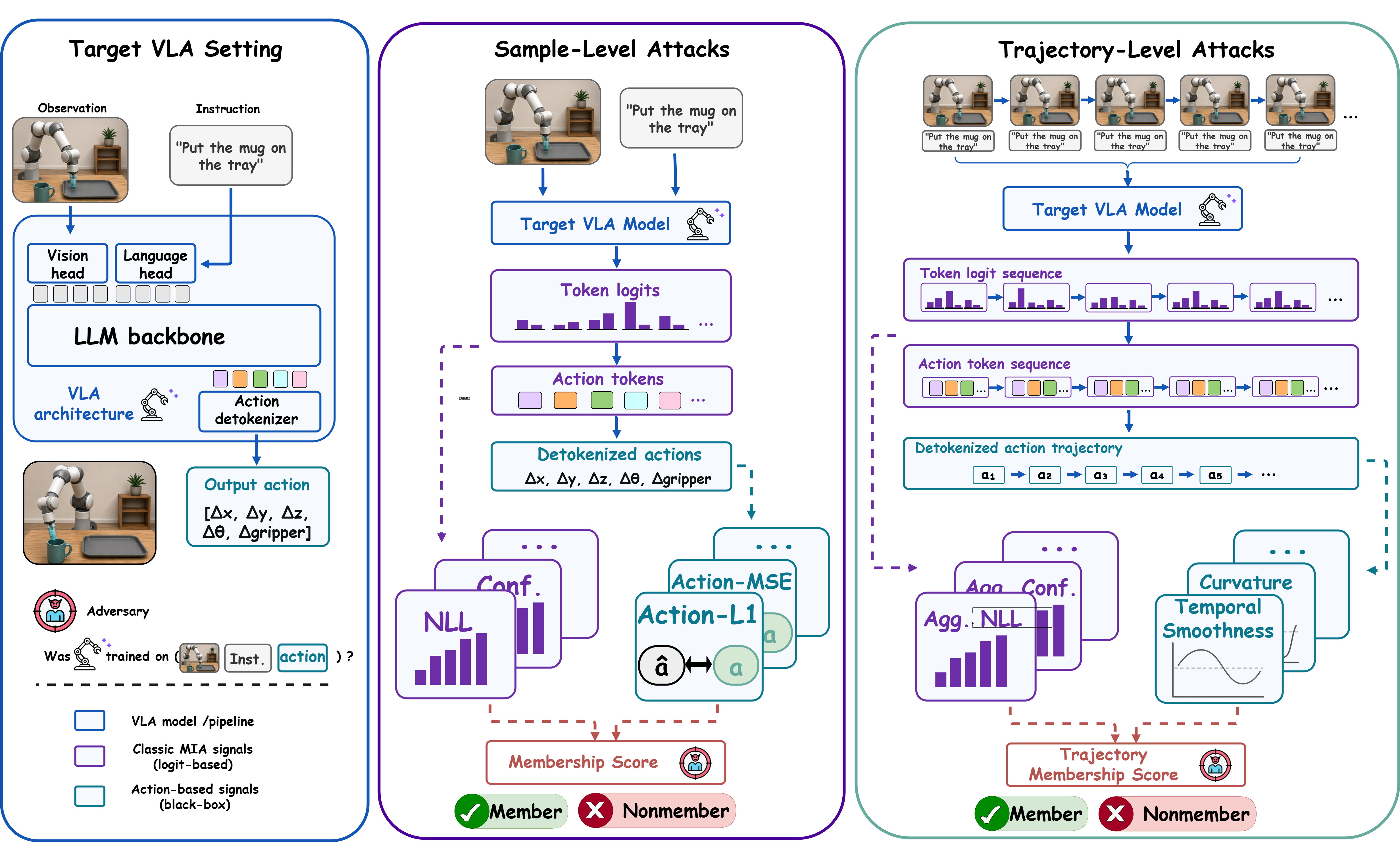}
    \vspace{-0.5em}
    \caption{
    Overview of MIA on VLA models. 
    We study sample-level attacks on individual transition samples and trajectory-level attacks on complete demonstrations. 
    The attacks exploit different signals, including token likelihood, generation confidence, action error, and temporal action dynamics, spanning both probability-based and black-box access settings.
    }
    \label{fig:vla_mia_overview}
    \vspace{-0.8em}
\end{figure*}

Through extensive experiments on representative LIBERO benchmarks~\cite{liu2023libero} and three representative VLA models~\cite{kim2024openvla,black2024pi_0,black2025pi_}, we demonstrate that these signals enable effective inference across a wide range of settings. At the sample level, classical likelihood-based attacks (e.g., NLL) achieve strong performance, reaching an average AUC of 0.8086 on OpenVLA, while even black-box action-only signals attain stronger performance, with Action-L1 and Action-MSE achieving average AUCs of 0.9233 and 0.9220, respectively. At the trajectory level, multiple methods achieve near-perfect inference. In particular, on OpenVLA, black-box temporal signals such as temporal smoothness and curvature reach average AUCs of 0.9989 and 0.9993 across datasets, respectively. 

Overall, our findings show that VLA models are highly vulnerable to membership inference, and that externally observable action behaviors can leak information about the training data in realistic embodied scenarios. This reveals a previously underexplored privacy risk in VLA systems and underscores the need for dedicated privacy evaluation and defenses for embodied AI.
\section{Related works}

\subsection{Attacks on Vision-Language-Action Models}

Vision-language-action (VLA) models map visual observations and language instructions to executable actions, making their outputs directly observable through physical or simulated behavior~\citep{rt12022arxiv,rt22023arxiv,kim2024openvla,black2024pi_0,black2025pi_}. 
This deployment interface creates an embodied attack surface beyond the textual or visual outputs typically studied in LLMs and VLMs.
Recent work has begun to study attacks against VLA systems, primarily focusing on integrity and control failures. 
Adversarial attacks manipulate visual observations or physical conditions to induce incorrect embodied decisions~\citep{wang2025advedm,cheng2024manipulation,wang2025adversarialvla,li2025attackvla, yan2025alignment}, while backdoor attacks implant trigger-conditioned behavioral deviations into VLA policies without substantially degrading clean-task performance~\citep{zhou2025badvla, xu2025tabvla}. 
Other studies further show that VLA robustness can depend on action tokenization and architecture-specific control interfaces~\citep{cheng2024manipulation,wang2025adversarialvla,li2025attackvla}.
These works highlight important safety and security risks in VLA deployment, but the privacy implications of VLA models remain largely overlooked. 
Our work addresses this complementary risk by studying membership privacy leakage in embodied data used to train VLA models.

\subsection{Membership Inference Attacks}

Membership inference attacks (MIAs) aim to determine whether a specific data record was used to train a target model~\cite{shokri2017membership}. 
They exploit behavioral differences between member and non-member data; for example, models often assign higher confidence or lower loss to samples seen during training. 
Early MIAs mainly studied image classification models~\cite{shokri2017membership,carlini2022membership,peng2024oslo,yeom2018privacy,nasr2019comprehensive}, while later work extended membership inference to foundation models, including large language models (LLMs)~\cite{shi2024detecting,duan2024membership, fu2024membership, hayes2026exploring, meeus2024did,chang2025context, mattern2023membership} and vision-language models (VLMs)~\cite{hu2025membership, li2024membership, ren2025self, liu2026lomia}.
However, MIAs on LLMs and VLMs are challenging because these models are typically trained on large-scale datasets for relatively few epochs, leading to weaker overfitting and less separable membership signals~\cite{duan2024membership,hu2025membership}. 
Recent studies therefore also consider broader notions of membership, such as dataset-level inference~\cite{maini2024llm, puerto2025scaling, tong2025how}, where evidence can be aggregated across multiple examples.

In contrast, membership inference against vision-language-action (VLA) models remains largely unexplored. 
VLA models are often fine-tuned for many epochs on relatively small embodied datasets, map observations and instructions to constrained action spaces, and expose structured actions as temporally correlated trajectories. 
These properties create a distinct attack surface for membership inference. 
Our work provides the first systematic study of this privacy risk, examining how membership information can leak through both internal model signals and externally observable action behaviors.
\section{Membership Inference on VLA Models}

\subsection{Problem Setup}

We consider a VLA model $f_\theta$ that maps a visual observation and a language instruction to an action. Let $\mathcal{D}_{\mathrm{train}}$ denote the training dataset of the target model. VLA training data is organized as a collection of embodied trajectories $\mathcal{D}_{\mathrm{train}} = \{\tau_i\}_{i=1}^{N}$.
Each trajectory $\tau_i$ consists of a sequence of transition samples:
\begin{equation}
    \tau_i = \{z_{i,t}\}_{t=1}^{T_i},
    \quad
    z_{i,t} = (o_{i,t}, x_i, a_{i,t}),
\end{equation}
where $o_{i,t}$ denotes the visual observation at time step $t$, $x_i$ denotes the language instruction, and $a_{i,t}$ denotes the corresponding action. Given an observation-instruction pair, the model predicts an action as
\begin{equation}
    \hat{a}_{i,t} = f_\theta(o_{i,t}, x_i).
\end{equation}

The trajectory-based structure of VLA training data naturally leads to two membership inference granularities: sample-level inference over individual transition samples and trajectory-level inference over complete embodied trajectories. We formalize both settings below.

\subsubsection{Sample-Level Membership Inference}

In sample-level MIA, the adversary is given a candidate transition sample $z=(o,x,a)$ and aims to determine whether this sample was used to train the target model. We define the ground-truth sample membership label as
\begin{equation}
    m_{\mathrm{sample}}(z)
    =
    \mathbf{1}\left[
        z \in \mathcal{D}_{\mathrm{train}}^{\mathrm{sample}}
    \right],
\end{equation}
where $m_{\mathrm{sample}}(z)=1$ indicates that $z$ is a member sample, and $m_{\mathrm{sample}}(z)=0$ indicates that it is a non-member sample. Given the target VLA model $f_\theta$ and the candidate sample $z$, the adversary constructs an inference function
\begin{equation}
    \mathcal{A}_{\mathrm{sample}}(f_\theta,z) \in \{0,1\},
\end{equation}
where output $1$ indicates that $z$ is predicted as a member. This setting captures membership leakage at the level of an individual embodied decision point, i.e., a single observation-instruction-action tuple.
\subsubsection{Trajectory-Level Membership Inference}

In trajectory-level MIA, the adversary is given a candidate trajectory $\tau=\{z_t\}_{t=1}^{T}$, where $z_t=(o_t,x,a_t)$, and aims to determine whether the trajectory should be considered a member. We define trajectory membership based on the fraction of member samples in the trajectory:
\begin{equation}
    m_{\mathrm{traj}}(\tau)
    =
    \mathbf{1}\left[
        \frac{1}{T}\sum_{t=1}^{T} m_{\mathrm{sample}}(z_t)
        \geq \rho
    \right],
\end{equation}
where $\rho \in (0,1]$ is the trajectory membership threshold. When $\rho=1$, a trajectory is considered a member only if all of its transition samples are members.

Given the target model $f_\theta$ and the candidate trajectory $\tau$, the adversary constructs an inference function
\begin{equation}
    \mathcal{A}_{\mathrm{traj}}(f_\theta,\tau) \in \{0,1\},
\end{equation}
where output $1$ indicates that $\tau$ is predicted as a member trajectory.

Trajectory-level MIA captures whether a full embodied interaction episode, rather than an individual transition, has sufficient membership evidence. This setting is particularly relevant for VLA models because embodied datasets are naturally organized as trajectories, and a full trajectory may encode sensitive information about user behavior, environments, or task workflows.

\subsection{Attack Methods}

We now introduce the MIA methods used in our study. 
Each attack computes a scalar membership score $s(\cdot)$ for a candidate sample or trajectory, where a higher score indicates stronger evidence of membership. 
The final prediction is obtained by thresholding the score:
\begin{equation}
    \mathcal{A}(f_\theta, q)
    =
    \mathbf{1}\left[s(q; f_\theta) \geq \gamma\right],
\end{equation}
where $q$ denotes either a transition sample $z$ or a trajectory $\tau$, and $\gamma$ is a decision threshold. Table~\ref{tab:attack_methods} summarizes the access requirements of the attacks considered in this work. We next describe each method in detail.

\begin{table*}[t]
\centering
\footnotesize
\setlength{\tabcolsep}{6pt}
\renewcommand{\arraystretch}{1.10}
\caption{Access requirements of VLA MIAs. \cmark/\xmark indicates whether the signal is used. ``Black-box'' means the attack only observes generated actions from the model.} 
\label{tab:attack_methods}
\begin{tabular}{@{}llccccc@{}}
\toprule
\textbf{Granularity} &
\textbf{Attack} &
\makecell[c]{\textbf{Original}\\\textbf{instruction}} &
\makecell[c]{\textbf{Ground-truth}\\\textbf{action}} &
\makecell[c]{\textbf{Token}\\\textbf{probabilities}} &
\makecell[c]{\textbf{Generated}\\\textbf{action}} &
\makecell[c]{\textbf{Black-box}} \\
\midrule

\multirow{5}{*}{Sample}
& NLL              & \cmark & \cmark & \cmark & \xmark & \xmark \\
& Conf             & \cmark & \xmark & \cmark & \cmark & \xmark \\
& Conf-Fix         & \xmark & \xmark & \cmark & \cmark & \xmark \\
& Action-$\ell_1$  & \cmark & \cmark & \xmark & \cmark & \cmark \\
& Action-MSE       & \cmark & \cmark & \xmark & \cmark & \cmark \\
\midrule

\multirow{2}{*}{Trajectory}
& Temp.-Smooth     & \cmark & \xmark & \xmark & \cmark & \cmark \\
& Temp.-Curv.      & \cmark & \xmark & \xmark & \cmark & \cmark \\

\bottomrule
\end{tabular}
\end{table*}

\subsubsection{Sample-Level Attacks}

For a transition sample $z=(o,x,a)$, we consider three types of membership scores. 

\paragraph{Action-token likelihood.}
Many VLA models represent actions as discrete action tokens. 
Let $a=(y_1,\ldots,y_L)$ denote the tokenized ground-truth action. 
We compute the mean log-likelihood of the ground-truth action tokens under teacher forcing:
\begin{equation}
    s_{\mathrm{nll}}(z; f_\theta)
    =
    \frac{1}{L}
    \sum_{j=1}^{L}
    \log p_\theta(y_j \mid o, x, y_{<j}).
\end{equation}
This attack requires access to both the ground-truth action tokens and the model's token probabilities. 
Since the model is optimized on training action tokens during fine-tuning, member samples are expected to receive higher likelihood.

\paragraph{Generation confidence.}
We next measure the model's confidence during autoregressive action generation. 
Let $\hat{y}_1,\ldots,\hat{y}_L$ be the generated action tokens. 
We average the maximum log-probability at each decoding step:
\begin{equation}
    s_{\mathrm{conf}}(z; f_\theta)
    =
    \frac{1}{L}
    \sum_{j=1}^{L}
    \log \max_{v \in \mathcal{V}}
    p_\theta(v \mid o, x, \hat{y}_{<j}).
\end{equation}

Unlike action-token likelihood, this attack does not require the ground-truth action, but it still requires access to token probabilities. 
We evaluate it with both the original instruction and a fixed generic prompt, e.g., ``What action should the robot take?'', to test whether confidence-based leakage remains when the original instruction context is unavailable.

\paragraph{Generated action error.}
Finally, we compare the generated continuous action $\hat{a}=f_\theta(o,x)$ with the ground-truth action $a$. 
We use the negative prediction error as the membership score:
\begin{equation}
    s_{\ell_1}(z; f_\theta)
    =
    -\|\hat{a}-a\|_1,
    \qquad
    s_{\mathrm{mse}}(z; f_\theta)
    =
    -\frac{1}{d}\|\hat{a}-a\|_2^2,
\end{equation}
where $d$ is the action dimension. 
These attacks require only the generated action and the candidate ground-truth action, without accessing token probabilities or other internal model signals. 
They therefore represent black-box attacks based on observable action outputs.

\subsubsection{Trajectory-Level Attacks}

For a candidate trajectory $\tau=\{z_t\}_{t=1}^{T}$, we consider two classes of trajectory-level attacks: aggregating sample-level evidence and exploiting temporal dynamics in generated actions.

\paragraph{Score aggregation.}
A natural trajectory-level attack aggregates membership evidence across all transition samples. 
Given a sample-level score $s(z_t; f_\theta)$, we define
\begin{equation}
    s_{\mathrm{agg}}(\tau; f_\theta)
    =
    \frac{1}{T}
    \sum_{t=1}^{T}
    s(z_t; f_\theta).
\end{equation}
This aggregation can be applied to any sample-level score introduced above. 
By combining evidence over multiple timesteps, trajectory-level aggregation can amplify membership signals that may be weak or noisy for individual samples.

\paragraph{Temporal black-box scores.}
Beyond per-step prediction quality, VLA models expose generated action sequences, which may reveal temporal behavioral patterns absent from standard LLM/VLM query settings. 
The intuition is that a model may reproduce training demonstrations more stably on member trajectories, leading to smoother and more consistent generated actions. 
We therefore define two black-box scores based only on generated actions.

Let $\hat{a}_t=f_\theta(o_t,x)$ denote the generated action at time step $t$. 
We first measure temporal smoothness as the negative average first-order action difference:
\begin{equation}
    s_{\mathrm{smooth}}(\tau; f_\theta)
    =
    -\frac{1}{T-1}
    \sum_{t=2}^{T}
    \|\hat{a}_t-\hat{a}_{t-1}\|_2.
\end{equation}
We further measure temporal curvature as the negative average second-order action difference:
\begin{equation}
    s_{\mathrm{curv}}(\tau; f_\theta)
    =
    -\frac{1}{T-2}
    \sum_{t=3}^{T}
    \|\hat{a}_t - 2\hat{a}_{t-1} + \hat{a}_{t-2}\|_2.
\end{equation}
Higher scores indicate smoother or lower-curvature generated trajectories, which may suggest stronger familiarity with the queried trajectory. 
Since these scores rely only on observable model outputs and do not require the ground-truth action sequence, they represent a particularly practical and concerning black-box attack setting.
\section{Experiments}

\subsection{Experimental Setup}

\paragraph{Models and Datasets.}
Our main experiments use OpenVLA~\cite{kim2024openvla} and $\pi_0$-fast~\cite{black2024pi_0}, which represent two widely used VLA architectures. 
We additionally report results on $\pi_{0.5}$~\cite{black2025pi_} in Appendix~\ref{app:pi05}, which represents a different architecture that augments the backbone with an action expert module.

We conduct experiments on four LIBERO benchmarks~\cite{liu2023libero}: Spatial, Object, Goal, and Long. 
For each dataset, we split the trajectories into two disjoint subsets: one half is used for model fine-tuning and treated as member data, while the other half is reserved as non-member data. 
Unless otherwise specified, we follow the official training configurations and hyperparameters from the original model implementations. 
Additional implementation details are provided in Appendix~\ref{app:implementation_details}.

\paragraph{Metrics.}
We evaluate attack performance using the area under the ROC curve (AUC) and true positive rate at low false positive rates (TPR@FPR). 
AUC measures the overall separability between member and non-member samples or trajectories. 
TPR@FPR measures the attack success rate under strict false-positive constraints, which is especially important in privacy auditing scenarios where false accusations should be rare~\cite{carlini2022membership}.

\subsection{Sample-Level MIA Results}

Table~\ref{tab:sample_level_results} reports sample-level MIA results across four LIBERO datasets and two VLA models. 
Overall, \textbf{both models exhibit strong sample-level membership leakage}. 
For $\pi_0$-fast, NLL is nearly perfect across all datasets, achieving an average AUC of $0.9998$ and an average TPR of $0.9543$ at $0.1\%$ FPR. 
This shows that ground-truth action likelihood is highly revealing when token probabilities are available. 
OpenVLA is also vulnerable to likelihood-based attacks, with NLL reaching an average AUC of $0.8086$ and up to $0.8906$ on the Object dataset.

The results further show that \textbf{membership leakage persists even without ground-truth action likelihood}. 
The generation-confidence attack achieves non-trivial performance, especially on OpenVLA, where it obtains an average AUC of $0.8421$ and reaches $0.8811$ on Long. 
However, Conf-FixPrompt performs close to random guessing. 
This suggests that confidence-based leakage depends heavily on the original instruction context; using only the image with a generic action prompt is often insufficient to recover a reliable membership signal.

Most notably, \textbf{black-box attacks are highly effective}. 
Action-L1 and Action-MSE achieve average AUCs of $0.9731$ and $0.9604$ on $\pi_0$-fast, and $0.9233$ and $0.9220$ on OpenVLA, respectively. 
They also remain strong in low-FPR regimes: on OpenVLA, Action-L1 and Action-MSE achieve average TPRs of $0.7018$ and $0.7111$ at $1\%$ FPR. 
These attacks do not require token probabilities or other internal model signals. 
This is particularly concerning because black-box MIAs against LLMs and VLMs are often limited~\cite{hu2025membership, liu2026lomia} when only generated outputs are available. In contrast, VLA outputs are structured, low-dimensional, and directly tied to action supervision, making observable actions themselves a strong black-box membership signal.

Together, these results show significant sample-level membership leakage in VLA models. Even externally observable actions provide strong membership evidence, highlighting a practical privacy risk for VLA systems.

\begin{table*}[t]
\centering
\scriptsize
\setlength{\tabcolsep}{3.5pt}
\renewcommand{\arraystretch}{0.9}
\caption{Sample-level membership inference results on four LIBERO datasets and two VLA models. For each model, we report AUC and TPR at FPR thresholds of 0.1\%, 1\%, and 5\%. Best results within each dataset-model block are bolded.}
\label{tab:sample_level_results}
\begin{tabular*}{\textwidth}{@{\extracolsep{\fill}}ll|cccc|cccc}
\toprule
\multirow{3}{*}{\textbf{Dataset}} &
\multirow{3}{*}{\textbf{Attack}} &
\multicolumn{4}{c|}{\textbf{$\pi_0$-fast}} &
\multicolumn{4}{c}{\textbf{OpenVLA}} \\
\cmidrule(lr){3-6}
\cmidrule(l){7-10}
& &
\multirow{2}{*}{\textbf{AUC}} &
\multicolumn{3}{c|}{\textbf{TPR@FPR}} &
\multirow{2}{*}{\textbf{AUC}} &
\multicolumn{3}{c}{\textbf{TPR@FPR}} \\
\cmidrule(lr){4-6}
\cmidrule(l){8-10}
& &
& \textbf{0.1\%} & \textbf{1\%} & \textbf{5\%}
& & \textbf{0.1\%} & \textbf{1\%} & \textbf{5\%} \\
\midrule

\multirow{5}{*}{Spatial}
& NLL
& \textbf{1.0000} & \textbf{0.9941} & \textbf{1.0000} & \textbf{1.0000}
& 0.8166 & 0.0452 & 0.1611 & 0.3620 \\
& Conf
& 0.7111 & 0.0509 & 0.1335 & 0.2872
& 0.8377 & 0.0426 & 0.1682 & 0.3818 \\
& Conf-FixPrompt
& 0.5151 & 0.0047 & 0.0169 & 0.0680
& 0.4996 & 0.0007 & 0.0095 & 0.0485 \\
& Action-L1
& 0.9801 & 0.4494 & 0.8020 & 0.9447
& \textbf{0.9233} & 0.6002 & 0.6906 & 0.7922 \\
& Action-MSE
& 0.9686 & 0.4530 & 0.8061 & 0.9433
& 0.9222 & \textbf{0.6063} & \textbf{0.7020} & \textbf{0.7989} \\
\midrule

\multirow{5}{*}{Goal}
& NLL
& \textbf{0.9999} & \textbf{0.9962} & \textbf{0.9998} & \textbf{1.0000}
& 0.7927 & 0.0114 & 0.0835 & 0.2755 \\
& Conf
& 0.7896 & 0.0362 & 0.1480 & 0.3582
& 0.8410 & 0.0494 & 0.1712 & 0.4229 \\
& Conf-FixPrompt
& 0.5618 & 0.0050 & 0.0301 & 0.0964
& 0.5006 & 0.0010 & 0.0082 & 0.0467 \\
& Action-L1
& 0.9854 & 0.2124 & 0.8722 & 0.9601
& \textbf{0.9207} & \textbf{0.5250} & 0.6990 & 0.7936 \\
& Action-MSE
& 0.9803 & 0.2457 & 0.8460 & 0.9533
& 0.9184 & 0.4486 & \textbf{0.7035} & \textbf{0.7980} \\
\midrule

\multirow{5}{*}{Object}
& NLL
& \textbf{0.9996} & \textbf{0.9392} & \textbf{0.9990} & \textbf{1.0000}
& 0.8906 & 0.1440 & 0.3165 & 0.5550 \\
& Conf
& 0.6817 & 0.0166 & 0.1091 & 0.2193
& 0.8086 & 0.0380 & 0.1596 & 0.3534 \\
& Conf-FixPrompt
& 0.5717 & 0.0054 & 0.0311 & 0.1012
& 0.5037 & 0.0006 & 0.0090 & 0.0524 \\
& Action-L1
& 0.9747 & 0.1925 & 0.5948 & 0.9163
& 0.9058 & \textbf{0.5687} & 0.6687 & 0.7562 \\
& Action-MSE
& 0.9672 & 0.1893 & 0.6228 & 0.9146
& \textbf{0.9064} & 0.5679 & \textbf{0.6792} & \textbf{0.7699} \\
\midrule

\multirow{5}{*}{Long}
& NLL
& \textbf{0.9995} & \textbf{0.8878} & \textbf{0.9915} & \textbf{0.9997}
& 0.7344 & 0.0092 & 0.0700 & 0.2147 \\
& Conf
& 0.5518 & 0.0086 & 0.0437 & 0.1185
& 0.8811 & 0.1068 & 0.3013 & 0.5294 \\
& Conf-FixPrompt
& 0.4958 & 0.0021 & 0.0150 & 0.0639
& 0.5007 & 0.0009 & 0.0087 & 0.0500 \\
& Action-L1
& 0.9520 & 0.3016 & 0.6378 & 0.8512
& \textbf{0.9435} & 0.6388 & 0.7490 & 0.8449 \\
& Action-MSE
& 0.9254 & 0.2508 & 0.6162 & 0.8262
& 0.9410 & \textbf{0.6494} & \textbf{0.7597} & \textbf{0.8511} \\
\bottomrule
\end{tabular*}
\end{table*}

\subsection{Trajectory-Level MIA Results}

Table~\ref{tab:trajectory_level_results} reports trajectory-level MIA results. 
Overall, \textbf{trajectory-level inference is substantially stronger than sample-level inference}, as aggregating evidence across a full demonstration amplifies membership signals. 
For $\pi_0$-fast, Agg.-NLL achieves perfect performance on all datasets, with an average AUC of $1.0000$ and TPR of $1.0000$ at all FPR thresholds. 
Action-based aggregation is also highly effective, with Action-L1 and Action-MSE reaching average AUCs of $0.9733$ and $0.9628$, respectively.
OpenVLA shows even stronger trajectory-level leakage. 
Agg.-NLL and Agg.-Conf achieve near-perfect performance across all datasets, with average AUCs of $0.9999$ and $1.0000$, respectively. 
Action-based aggregation is also nearly perfect: Action-L1 achieves an average AUC of $1.0000$, while Action-MSE achieves $0.9998$. 
These results indicate that membership signals that are already visible at individual timesteps become much more separable when accumulated over an entire trajectory.

Most importantly, \textbf{temporal black-box attacks are highly effective}. For OpenVLA,
Temp.-Smooth and Temp.-Curve achieve average AUCs of $0.9989$ and $0.9993$, with average TPRs of $0.9725$ and $0.9775$ at $0.1\%$ FPR, respectively. 
These attacks rely only on generated action sequences and do not require ground-truth actions, token probabilities, or internal model access. 
This finding is particularly concerning because it shows that VLA models can leak membership through temporal action dynamics alone, a signal that is unique to embodied action outputs. The strong results from aggregation-based and OpenVLA temporal attacks show that trajectory-level VLA outputs expose a rich and practical attack surface for membership inference.

\begin{table*}[t]
\centering
\scriptsize
\setlength{\tabcolsep}{3.2pt}
\renewcommand{\arraystretch}{0.9}
\caption{Trajectory-level membership inference results on four LIBERO datasets and two VLA models. For each model, we report AUC and TPR at FPR thresholds of 0.1\%, 1\%, and 5\%. Best results within each dataset-model block are bolded.}
\label{tab:trajectory_level_results}
\begin{tabular*}{\textwidth}{@{\extracolsep{\fill}}ll|cccc|cccc}
\toprule
\multirow{3}{*}{\textbf{Dataset}} &
\multirow{3}{*}{\textbf{Attack}} &
\multicolumn{4}{c|}{\textbf{$\pi_0$-fast}} &
\multicolumn{4}{c}{\textbf{OpenVLA}} \\
\cmidrule(lr){3-6}
\cmidrule(l){7-10}
& &
\multirow{2}{*}{\textbf{AUC}} &
\multicolumn{3}{c|}{\textbf{TPR@FPR}} &
\multirow{2}{*}{\textbf{AUC}} &
\multicolumn{3}{c}{\textbf{TPR@FPR}} \\
\cmidrule(lr){4-6}
\cmidrule(l){8-10}
& &
& \textbf{0.1\%} & \textbf{1\%} & \textbf{5\%}
& & \textbf{0.1\%} & \textbf{1\%} & \textbf{5\%} \\
\midrule

\multirow{7}{*}{Spatial}
& Agg.-NLL
& \textbf{1.0000} & \textbf{1.00} & \textbf{1.00} & \textbf{1.00}
& \textbf{1.0000} & \textbf{1.00} & \textbf{1.00} & \textbf{1.00} \\
& Agg.-Conf
& 0.9016 & 0.70 & 0.73 & 0.76
& \textbf{1.0000} & \textbf{1.00} & \textbf{1.00} & \textbf{1.00} \\
& Agg.-Conf-Fix
& 0.5673 & 0.10 & 0.11 & 0.19
& 0.4824 & 0.01 & 0.01 & 0.08 \\
& Temp.-Smooth
& 0.5504 & 0.00 & 0.04 & 0.10
& 0.9988 & 0.97 & 0.97 & 0.99 \\
& Temp.-Curve
& 0.6891 & 0.03 & 0.07 & 0.22
& 0.9992 & 0.98 & 0.98 & \textbf{1.00} \\
& Action-L1
& 0.9804 & 0.98 & 0.98 & 0.98
& \textbf{1.0000} & \textbf{1.00} & \textbf{1.00} & \textbf{1.00} \\
& Action-MSE
& 0.9616 & 0.96 & 0.96 & 0.96
& 0.9998 & 0.98 & \textbf{1.00} & \textbf{1.00} \\
\midrule

\multirow{7}{*}{Goal}
& Agg.-NLL
& \textbf{1.0000} & \textbf{1.00} & \textbf{1.00} & \textbf{1.00}
& \textbf{1.0000} & \textbf{1.00} & \textbf{1.00} & \textbf{1.00} \\
& Agg.-Conf
& 0.9971 & 0.95 & 0.96 & 0.98
& \textbf{1.0000} & \textbf{1.00} & \textbf{1.00} & \textbf{1.00} \\
& Agg.-Conf-Fix
& 0.7252 & 0.05 & 0.16 & 0.32
& 0.4774 & 0.00 & 0.04 & 0.07 \\
& Temp.-Smooth
& 0.6583 & 0.09 & 0.09 & 0.16
& 0.9968 & 0.92 & 0.94 & 0.98 \\
& Temp.-Curve
& 0.7475 & 0.15 & 0.15 & 0.21
& 0.9979 & 0.93 & 0.97 & 0.99 \\
& Action-L1
& 0.9615 & 0.96 & 0.96 & 0.96
& \textbf{1.0000} & \textbf{1.00} & \textbf{1.00} & \textbf{1.00} \\
& Action-MSE
& 0.9607 & 0.93 & 0.95 & 0.96
& 0.9993 & 0.98 & 0.99 & 0.99 \\
\midrule

\multirow{7}{*}{Object}
& Agg.-NLL
& \textbf{1.0000} & \textbf{1.00} & \textbf{1.00} & \textbf{1.00}
& \textbf{1.0000} & \textbf{1.00} & \textbf{1.00} & \textbf{1.00} \\
& Agg.-Conf
& 0.9533 & 0.74 & 0.77 & 0.86
& \textbf{1.0000} & \textbf{1.00} & \textbf{1.00} & \textbf{1.00} \\
& Agg.-Conf-Fix
& 0.7226 & 0.28 & 0.31 & 0.44
& 0.5699 & 0.02 & 0.02 & 0.11 \\
& Temp.-Smooth
& 0.6116 & 0.00 & 0.01 & 0.08
& \textbf{1.0000} & \textbf{1.00} & \textbf{1.00} & \textbf{1.00} \\
& Temp.-Curve
& 0.7729 & 0.14 & 0.15 & 0.31
& \textbf{1.0000} & \textbf{1.00} & \textbf{1.00} & \textbf{1.00} \\
& Action-L1
& 0.9903 & 0.99 & 0.99 & 0.99
& \textbf{1.0000} & \textbf{1.00} & \textbf{1.00} & \textbf{1.00} \\
& Action-MSE
& 0.9902 & 0.99 & 0.99 & 0.99
& \textbf{1.0000} & \textbf{1.00} & \textbf{1.00} & \textbf{1.00} \\
\midrule

\multirow{7}{*}{Long}
& Agg.-NLL
& \textbf{1.0000} & \textbf{1.00} & \textbf{1.00} & \textbf{1.00}
& 0.9996 & 0.98 & 0.98 & \textbf{1.00} \\
& Agg.-Conf
& 0.6462 & 0.10 & 0.12 & 0.19
& \textbf{1.0000} & \textbf{1.00} & \textbf{1.00} & \textbf{1.00} \\
& Agg.-Conf-Fix
& 0.5489 & 0.03 & 0.03 & 0.10
& 0.4839 & 0.00 & 0.00 & 0.10 \\
& Temp.-Smooth
& 0.6100 & 0.00 & 0.00 & 0.08
& \textbf{1.0000} & \textbf{1.00} & \textbf{1.00} & \textbf{1.00} \\
& Temp.-Curve
& 0.6716 & 0.01 & 0.02 & 0.15
& \textbf{1.0000} & \textbf{1.00} & \textbf{1.00} & \textbf{1.00} \\
& Action-L1
& 0.9612 & 0.96 & 0.96 & 0.96
& \textbf{1.0000} & \textbf{1.00} & \textbf{1.00} & \textbf{1.00} \\
& Action-MSE
& 0.9387 & 0.93 & 0.93 & 0.93
& \textbf{1.0000} & \textbf{1.00} & \textbf{1.00} & \textbf{1.00} \\
\bottomrule
\end{tabular*}
\end{table*}

\subsection{Ablation Studies}

\subsubsection{Effect of Action Output Space}

\begin{figure}[t]
    \centering
    \begin{subfigure}[t]{0.48\textwidth}
        \centering
        \includegraphics[width=\linewidth]{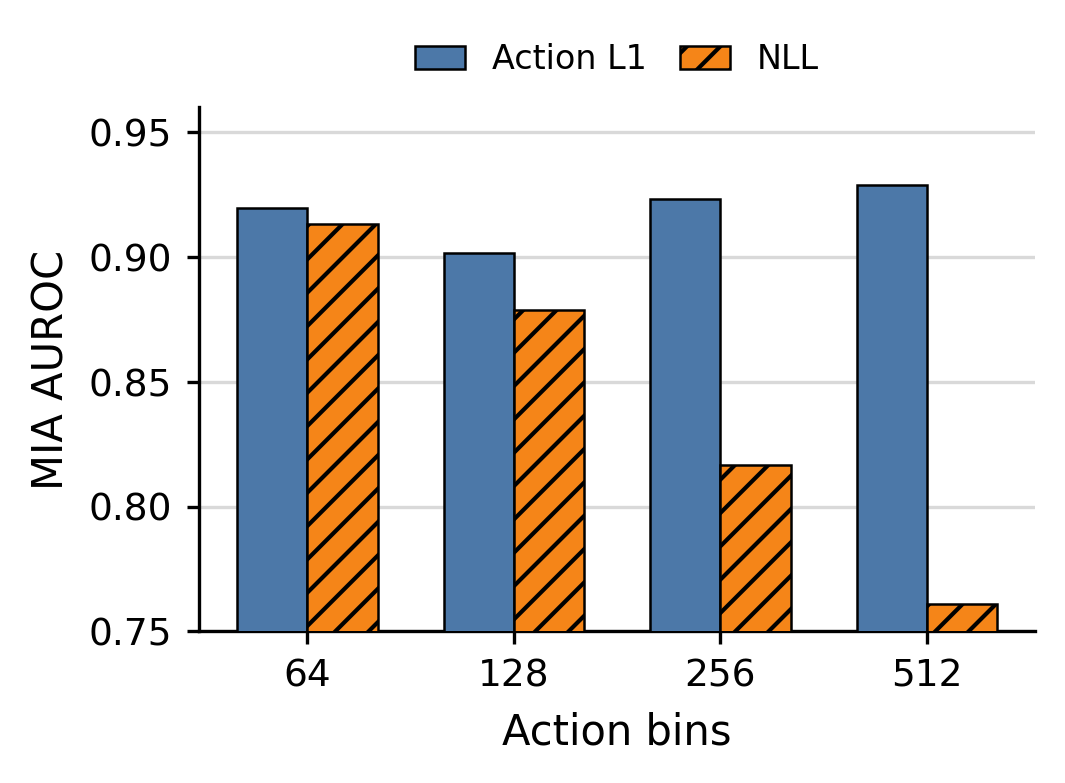}
        \caption{Action discretization.}
        \label{fig:bins_ablation}
    \end{subfigure}
    \hfill
    \begin{subfigure}[t]{0.48\textwidth}
        \centering
        \includegraphics[width=\linewidth]{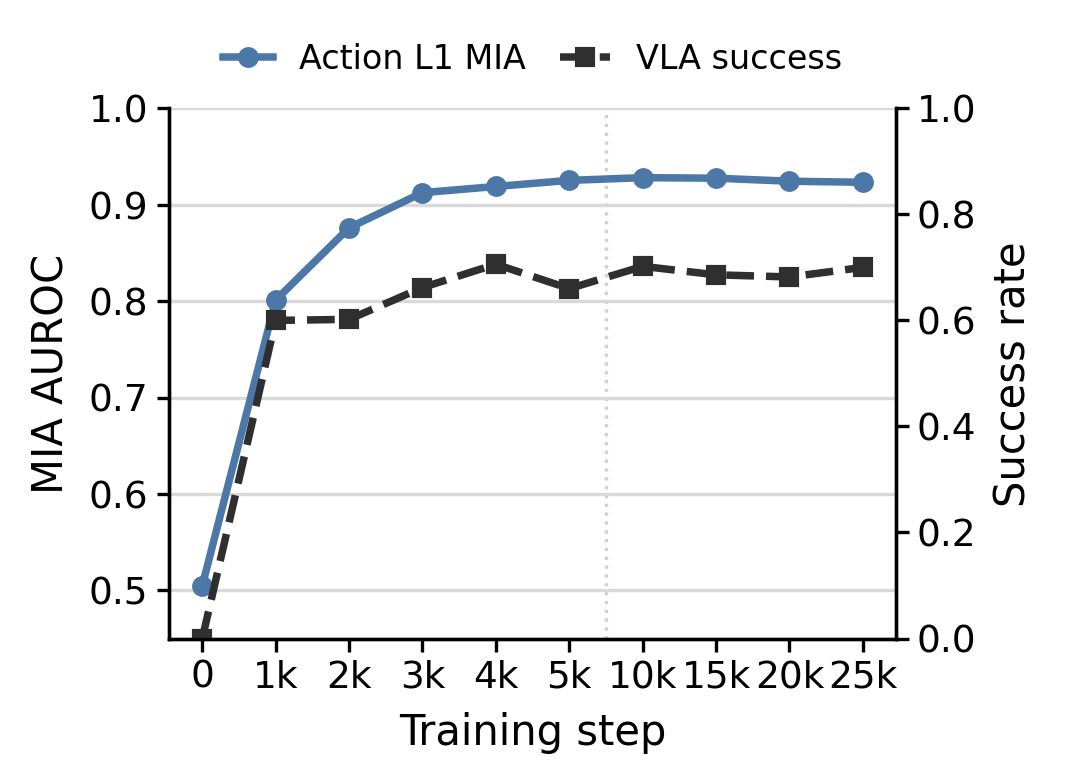}
        \caption{Fine-tuning steps.}
        \label{fig:steps_ablation}
    \end{subfigure}
    \caption{Ablation studies on OpenVLA using LIBERO-Spatial. 
    (a) Sample-level MIA AUC under different action discretization bin sizes. 
    (b) Task success rate and sample-level Action-L1 MIA AUC across fine-tuning steps.}
    \label{fig:ablation_studies}
    \vspace{-1.0em}
\end{figure}

We study how the VLA action output space affects membership leakage by varying the number of action bins in OpenVLA. 
As shown in Figure~\ref{fig:bins_ablation}, NLL-based MIA is sensitive to action discretization: increasing the number of bins from $64$ to $512$ reduces NLL AUC from $0.9133$ to $0.7608$. 
This suggests that a smaller and coarser action-token space amplifies likelihood differences between member and non-member samples, making classical loss-based membership signals more separable.

However, this trend does not hold for action-level attacks. 
Action-L1 remains consistently strong across different bin sizes, with AUCs between $0.9017$ and $0.9290$. 
This indicates that token-level likelihood does not fully capture the membership leakage exposed by VLA models. 
Although discretization changes the token prediction space and weakens the NLL signal, the predicted tokens are ultimately mapped back to executable actions. 
During this mapping, small token-level differences can still lead to systematic action-level differences between member and non-member samples.

This result highlights a VLA-specific privacy risk: membership leakage can persist in the final action behavior even when classical likelihood-based signals become weaker. Thus, evaluating VLA privacy only through loss or token likelihood can underestimate leakage, as the executable action space itself provides an additional and robust attack surface.

\subsubsection{Effect of Training Steps}

We study how membership leakage evolves during VLA fine-tuning by evaluating OpenVLA checkpoints at different training steps on LIBERO-Spatial. As shown in Figure~\ref{fig:steps_ablation}, membership leakage emerges early: Action-L1 AUC increases from $0.5049$ before fine-tuning to $0.8011$ after only $1$k steps, and further rises to $0.9125$ by $3$k steps. After $5$k steps, the leakage remains consistently high, with AUCs around $0.92$--$0.93$ through the final checkpoint.

Importantly, membership leakage closely tracks task performance during fine-tuning. The task success rate increases from $60.0\%$ at $1$k steps to around $70\%$ in later checkpoints, while Action-L1 AUC rises and then stabilizes over the same training period. This indicates that the same training process that improves the VLA policy also makes training samples more identifiable. This coupling between utility and leakage suggests that simple early stopping may be insufficient as a defense. Although stopping earlier could reduce membership leakage, it would also reduce task success, creating a direct privacy-utility trade-off. This is particularly concerning for VLA models, where fine-tuning on relatively small embodied datasets often requires repeated exposure to the same demonstrations to achieve strong downstream performance~\cite{kim2024openvla}.
\section{Potential Mitigations}

\begin{wrapfigure}{r}{0.43\textwidth}
    \vspace{-1.0em}
    \centering
    \includegraphics[width=0.41\textwidth]{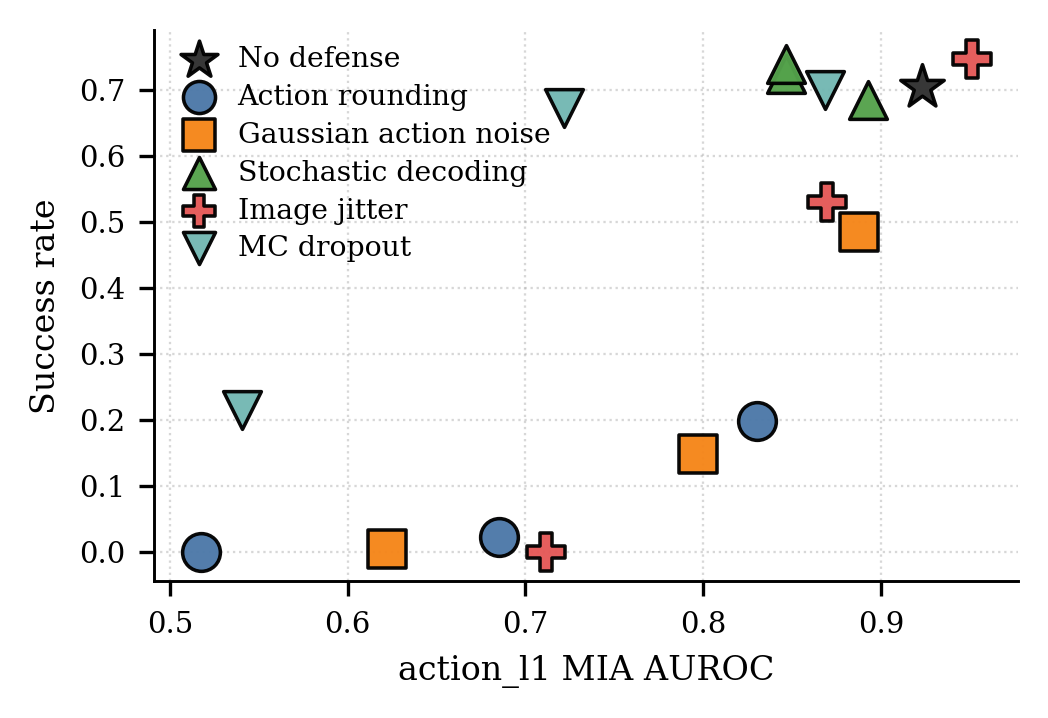}
    \vspace{-0.8em}
    \caption{Privacy-utility trade-off of mitigations on OpenVLA with LIBERO-Spatial.}
    \label{fig:mitigation_tradeoff}
    \vspace{-1.0em}
\end{wrapfigure}

We evaluate mitigations against action-based MIAs, which remain feasible under black-box access and thus form a practical leakage channel in deployed VLA systems. We consider five defenses: Gaussian action noise, action rounding, stochastic decoding, image jitter, and MC dropout, each with multiple settings; details are in Appendix~\ref{app:mitigation}. These methods aim to reduce fine-grained action memorization signals.

Figure~\ref{fig:mitigation_tradeoff} shows a clear privacy-utility trade-off. 
Output perturbations such as Gaussian noise and action rounding reduce leakage but sharply degrade task performance, suggesting that the fine-grained action information exploited by MIAs is also essential for precise control. 
Input-side perturbations such as image jitter show a similar pattern, where stronger perturbation lowers MIA AUC but can collapse task success. 
MC dropout achieves a stronger trade-off, reducing Action-L1 AUC to $0.7216$ while preserving $67.2\%$ success, but more aggressive dropout still harms utility. 
Stochastic decoding is lighter-weight, reducing AUC to $0.8467$ while preserving $72$--$74\%$ success, yet substantial leakage remains. 
Overall, these results show that mitigating VLA membership leakage is non-trivial: effective defenses must reduce action-level memorization while preserving the precision required for embodied control, motivating VLA-specific privacy defenses.
\section{Conclusion}

We present the first systematic study of membership inference attacks against vision-language-action (VLA) models. We formalize sample-level and trajectory-level membership inference and evaluate attacks using both conventional signals, such as likelihood and confidence, and VLA-specific signals, such as action errors and temporal dynamics. Our results show that VLA models expose substantial membership leakage, with black-box attacks using only generated actions achieving strong performance. This highlights exposed action outputs as a practical privacy risk in deployed embodied AI systems. Our mitigation study further shows that reducing leakage without harming task performance is non-trivial, underscoring the need for VLA-specific privacy defenses.

\paragraph{Limitations.}
Our study focuses on representative VLA models and benchmarks, so results may vary across robot platforms, action spaces, datasets, and deployments. Due to the high cost of pretraining-scale experiments, we study fine-tuned VLA models. Our mitigation results are preliminary, leaving stronger VLA-specific defenses as future work.

\paragraph{Broader Impact.}
This work reveals an underexplored privacy risk in embodied AI. As VLA models are trained on data from homes, workplaces, hospitals, and other sensitive environments, membership leakage may expose whether specific users, routines, or demonstrations were included in training. Meanwhile, membership inference can also support responsible auditing of data provenance and unauthorized dataset use. We hope this work encourages privacy-aware evaluation, data governance, and defense design for future VLA systems.

\bibliographystyle{plain}
\bibliography{neurips_2026}

\appendix
\section{Appendix}

\subsection{Additional Implementation Details}
\label{app:implementation_details}

\paragraph{Fine-tuning setup.}
Unless otherwise specified, we follow the official training configurations of the original model implementations, while using half of each LIBERO dataset for fine-tuning and reserving the other half as non-member data; the number of training steps is correspondingly reduced to half of the official configuration. 
For OpenVLA, we use LoRA with rank $32$, learning rate $5\times 10^{-4}$, image augmentation enabled, and a shuffle buffer size of $100$k, trained on $4$ A100 GPUs with per-GPU batch size $16$ and gradient accumulation $2$ for an effective batch size of $128$. 
For $\pi_0$-FAST, we apply LoRA on the PaliGemma backbone with rank $16$ and learning rate $5\times 10^{-5}$, trained on $2$ A100 GPUs with effective batch size $32$. 
For $\pi_{0.5}$, we apply LoRA on both the PaliGemma backbone (rank $16$) and the action expert (rank $32$), with learning rate $5\times 10^{-5}$, trained on $2$ A100 GPUs with effective batch size $32$. 
All effective batch sizes match the official settings of the corresponding models.

\begin{table}[h]
\centering
\small
\setlength{\tabcolsep}{5pt}
\renewcommand{\arraystretch}{1.05}
\caption{Fine-tuning setup for OpenVLA on LIBERO datasets. We use half of each dataset for training, reduce the number of training steps accordingly, and keep the official effective batch size.}
\label{tab:finetuning_setup}
\begin{tabular}{lccc}
\toprule
\textbf{Dataset} & \textbf{Train Traj.} & \textbf{Steps}  & \textbf{Time} \\
\midrule
LIBERO-Spatial & 216 & 25k & $\sim$17h \\
LIBERO-Object  & 226 & 25k &  $\sim$17h \\
LIBERO-Goal    & 214 & 30k &  $\sim$21h \\
LIBERO-Long    & 190 & 40k &  $\sim$28h \\
\bottomrule
\end{tabular}
\end{table}

\paragraph{MIA evaluation.}
For sample-level evaluation, we randomly sample $10{,}000$ member transition samples and $10{,}000$ non-member transition samples from the training and reserved splits, respectively. 
For trajectory-level evaluation, we sample $100$ member trajectories and $100$ non-member trajectories. 

\subsection{Additional Results on $\pi_{0.5}$}
\label{app:pi05}

To evaluate whether membership leakage persists across different VLA architectures, we additionally report results on $\pi_{0.5}$, which augments the backbone with an action expert module. 
Since $\pi_{0.5}$ uses a flow-matching-based action generation mechanism, we adapt the likelihood-based attack by using Flow-Loss as the corresponding training-loss signal. 
We also evaluate action-based and trajectory-level attacks following the same protocol as in the main experiments.

Table~\ref{tab:pi05_mia_results} shows that $\pi_{0.5}$ also exhibits strong membership leakage. 
At the sample level, Action-L1 achieves high AUCs across all four datasets, ranging from $0.9009$ to $0.9481$. At the trajectory level, leakage becomes nearly perfect. 
Action-L1 and Agg.-Flow-Loss achieve AUCs close to or equal to $1.0$ across all datasets, demonstrating that trajectory-level aggregation strongly amplifies membership signals. Overall, these results confirm that membership leakage is not specific to a single VLA architecture.

\begin{table*}[t]
\centering
\scriptsize
\setlength{\tabcolsep}{5pt}
\renewcommand{\arraystretch}{1.08}
\caption{MIA results for $\pi_{0.5}$ across four LIBERO datasets. Best result within each suite-level block is bolded.}
\label{tab:pi05_mia_results}
\begin{tabular*}{\textwidth}{@{\extracolsep{\fill}}lll|cccc}
\toprule
\multirow{2}{*}{\textbf{Dataset}} &
\multirow{2}{*}{\textbf{Setting}} &
\multirow{2}{*}{\textbf{Attack Method}} &
\multirow{2}{*}{\textbf{AUC} $\uparrow$} &
\multicolumn{3}{c}{\textbf{TPR@FPR} $\uparrow$} \\
& & & & \textbf{FPR=0.1\%} & \textbf{FPR=1\%} & \textbf{FPR=5\%} \\
\midrule

\multirow{9}{*}{Spatial}
& \multirow{4}{*}{Sample}
& Action-L1                    & \textbf{0.9453} & 0.0631          & \textbf{0.3741} & \textbf{0.7468} \\
& & Action-MSE                 & 0.9115          & \textbf{0.0882} & 0.4267          & 0.7226          \\
& & Flow-Loss                  & 0.9041          & 0.0600          & 0.3004          & 0.5664          \\
& & Flow-Loss-FixPrompt        & 0.5212          & 0.0006          & 0.0077          & 0.0509          \\
\cmidrule(l){2-7}
& \multirow{5}{*}{Trajectory}
& Action-L1                    & \textbf{1.0000} & \textbf{1.0000} & \textbf{1.0000} & \textbf{1.0000} \\
& & Action-MSE                 & 0.9998          & 0.9900          & 0.9900          & \textbf{1.0000} \\
& & Agg.-Flow-Loss             & \textbf{1.0000} & \textbf{1.0000} & \textbf{1.0000} & \textbf{1.0000} \\
& & Temporal-Smoothness        & 0.5619          & 0.0100          & 0.0300          & 0.0600          \\
& & Temporal-Curvature         & 0.6536          & 0.0800          & 0.0900          & 0.1300          \\
\midrule

\multirow{9}{*}{Goal}
& \multirow{4}{*}{Sample}
& Action-L1                    & \textbf{0.9481} & 0.1062          & 0.4766          & \textbf{0.7641} \\
& & Action-MSE                 & 0.9064          & \textbf{0.1230} & \textbf{0.4972} & 0.7286          \\
& & Flow-Loss                  & 0.8914          & 0.0841          & 0.3023          & 0.5287          \\
& & Flow-Loss-FixPrompt        & 0.5260          & 0.0001          & 0.0101          & 0.0539          \\
\cmidrule(l){2-7}
& \multirow{5}{*}{Trajectory}
& Action-L1                    & \textbf{1.0000} & \textbf{1.0000} & \textbf{1.0000} & \textbf{1.0000} \\
& & Action-MSE                 & 0.9977          & 0.7900          & 0.9800          & \textbf{1.0000} \\
& & Agg.-Flow-Loss             & \textbf{1.0000} & \textbf{1.0000} & \textbf{1.0000} & \textbf{1.0000} \\
& & Temporal-Smoothness        & 0.5295          & 0.0300          & 0.0400          & 0.1200          \\
& & Temporal-Curvature         & 0.5976          & 0.0900          & 0.1000          & 0.1300          \\
\midrule

\multirow{9}{*}{Object}
& \multirow{4}{*}{Sample}
& Action-L1                    & \textbf{0.9258} & 0.0236          & 0.2301          & 0.5671          \\
& & Action-MSE                 & 0.9025          & \textbf{0.0506} & \textbf{0.2455} & \textbf{0.6046} \\
& & Flow-Loss                  & 0.8900          & 0.0371          & 0.2088          & 0.4684          \\
& & Flow-Loss-FixPrompt        & 0.4936          & 0.0011          & 0.0100          & 0.0456          \\
\cmidrule(l){2-7}
& \multirow{5}{*}{Trajectory}
& Action-L1                    & \textbf{1.0000} & \textbf{1.0000} & \textbf{1.0000} & \textbf{1.0000} \\
& & Action-MSE                 & \textbf{1.0000} & \textbf{1.0000} & \textbf{1.0000} & \textbf{1.0000} \\
& & Agg.-Flow-Loss             & \textbf{1.0000} & \textbf{1.0000} & \textbf{1.0000} & \textbf{1.0000} \\
& & Temporal-Smoothness        & 0.6248          & 0.0200          & 0.0200          & 0.1500          \\
& & Temporal-Curvature         & 0.6982          & 0.0900          & 0.0900          & 0.2000          \\
\midrule

\multirow{9}{*}{Long}
& \multirow{4}{*}{Sample}
& Action-L1                    & \textbf{0.9009} & 0.0468          & \textbf{0.2381} & \textbf{0.5806} \\
& & Action-MSE                 & 0.8618          & \textbf{0.0549} & 0.2314          & 0.5798          \\
& & Flow-Loss                  & 0.8468          & 0.0388          & 0.1994          & 0.4141          \\
& & Flow-Loss-FixPrompt        & 0.4858          & 0.0018          & 0.0127          & 0.0572          \\
\cmidrule(l){2-7}
& \multirow{5}{*}{Trajectory}
& Action-L1                    & 0.9996          & \textbf{0.9900} & \textbf{0.9900} & \textbf{1.0000} \\
& & Action-MSE                 & 0.9899          & 0.7200          & 0.7800          & 0.9900          \\
& & Agg.-Flow-Loss             & \textbf{0.9998} & \textbf{0.9900} & \textbf{0.9900} & \textbf{1.0000} \\
& & Temporal-Smoothness        & 0.5804          & 0.0000          & 0.0100          & 0.0600          \\
& & Temporal-Curvature         & 0.6203          & 0.0000          & 0.0100          & 0.1200          \\
\bottomrule
\end{tabular*}
\end{table*}

\subsection{Mitigation Details}
\label{app:mitigation}

We evaluate five mitigations against action-based MIAs, as detailed below. 

\paragraph{Gaussian action noise.}
After the model generates a continuous action vector $a$, we add Gaussian noise to the motion dimensions:
\begin{equation}
    a_{\mathrm{def}} = a + \epsilon,
    \qquad
    \epsilon \sim \mathcal{N}(0, \sigma^2 I).
\end{equation}
The gripper dimension is left unchanged. 
We evaluate $\sigma \in \{0.10, 0.20, 0.50\}$. 
This defense aims to reduce fine-grained action matching between generated actions and member samples.

\paragraph{Action rounding.}
We quantize each generated action dimension to a fixed grid:
\begin{equation}
    a_{\mathrm{def}}
    =
    \mathrm{round}\left(\frac{a}{\delta}\right)\cdot \delta.
\end{equation}
As with Gaussian noise, we apply rounding only to motion dimensions and leave the gripper dimension unchanged. 
We evaluate $\delta \in \{0.50, 1.00, 2.00\}$. 
This defense removes fine-grained continuous action information that may be exploited by action-distance attacks.

\paragraph{Stochastic decoding.}
Instead of greedy decoding for action tokens, we sample action tokens with temperature scaling:
\begin{equation}
    p_T(y_i \mid x, y_{<i})
    \propto
    p(y_i \mid x, y_{<i})^{1/T}.
\end{equation}
A larger temperature flattens the output distribution and makes generation less deterministic. 
We keep nucleus sampling fixed with top-$p=0.95$ and evaluate $T \in \{1.10, 2.00, 3.00\}$. 
This defense aims to reduce deterministic behavior on training samples during action-token generation.

\paragraph{Image jitter.}
We apply randomized image perturbations at inference time before passing the image to the VLA processor. 
Let $\mathcal{T}_{\mathrm{img}}$ denote a stochastic image transformation. 
The defended action is generated as
\begin{equation}
    a_{\mathrm{def}}
    =
    f_\theta(\mathcal{T}_{\mathrm{img}}(o), x).
\end{equation}
The transformation includes random cropping followed by resizing to the original resolution, brightness and contrast perturbations, and Gaussian pixel noise. 
We evaluate three strengths: light, medium, and strong jitter. 
This defense tests whether perturbing the visual input can reduce action-level membership signals.

\paragraph{Monte Carlo (MC) dropout.}
We enable dropout during inference while keeping the same trained checkpoint. 
Let $m$ denote a random dropout mask sampled at inference time. 
The defended action is generated as
\begin{equation}
    a_{\mathrm{def}}
    =
    f_{\theta,m}(o,x).
\end{equation}
We set dropout modules to train mode while keeping the rest of the model in evaluation mode. 
We evaluate dropout probabilities $p \in \{0.10, 0.20, 0.40\}$. 
This defense injects stochasticity into hidden activations to reduce deterministic action behavior on member samples.

Table~\ref{tab:mitigation_settings} summarizes the mitigation settings used in our experiments.

\begin{table}[htbp]
\centering
\small
\setlength{\tabcolsep}{5pt}
\renewcommand{\arraystretch}{1.08}
\caption{Mitigations evaluated on OpenVLA.}
\label{tab:mitigation_settings}
\begin{tabular}{ll}
\toprule
\textbf{Defense} & \textbf{Settings} \\
\midrule
Action rounding 
& $\delta \in \{0.50, 1.00, 2.00\}$ \\
Gaussian action noise 
& $\sigma \in \{0.10, 0.20, 0.50\}$ \\
Stochastic decoding 
& $T \in \{1.10, 2.00, 3.00\}$, top-$p=0.95$ \\
Image jitter 
& light / medium / strong perturbation \\
MC dropout 
& $p \in \{0.10, 0.20, 0.40\}$ \\
\bottomrule
\end{tabular}
\end{table}

For image jitter, the light, medium, and strong settings use minimum crop area fractions of $0.75$, $0.60$, and $0.45$, brightness and contrast ranges of $0.20$, $0.35$, and $0.50$, and Gaussian pixel-noise standard deviations of $0.05$, $0.10$, and $0.20$, respectively. 
For MC dropout, the historical setting names correspond to actual dropout probabilities of $0.10$, $0.20$, and $0.40$.

\end{document}